\documentclass[a4paper, british]{article}

\usepackage[margin=3cm]{geometry}

\usepackage[utf8]{inputenc}
\usepackage[T1]{fontenc}

\usepackage[german, main=british]{babel}
\usepackage{csquotes}

\usepackage{amsmath, amssymb, xcolor, booktabs}
\usepackage{tikz}
\graphicspath{{pics/}}

\usepackage{multicol}

\usepackage[normalem]{ulem}
\let\simplestrike\sout
\renewcommand{\sout}[1]{\textcolor{purple}{\simplestrike{#1}}}

\usepackage{hyperref}
\usepackage[style=phys, backend=biber, eprint]{biblatex}
\providecommand{\IC}{ {\ensuremath{\mathrm{IC}}} }
\providecommand{\PR}{ {\ensuremath{\mathrm{PR}}} }

\providecommand{\wat}{{\ensuremath{\!H_{\!2}\!O}}}
\providecommand{\ato}[1][i]{%
	{\ensuremath{\!A^{0}_{#1}}}%
}
\providecommand{\cat}{ {\ensuremath{\!A^{\!+}}} }
\providecommand{\ion}[1][i]{%
	{\ensuremath{\cat \gets\, {\ato[#1]}}}%
}
\providecommand{\rec}[1][i]{%
	{\ensuremath{\cat \to\,   {\ato[#1]}}}%
}

\providecommand{\iep}{ {\ensuremath{\!i,\epsilon}} }
\providecommand{\jep}{ {\ensuremath{\!j,\epsilon}} }

\providecommand{\dist}{ {\ensuremath{r}} }

\title{
	Water-assisted electron capture exceeds photo\-recombination in biological conditions}
\author{
	Axel Molle\textsuperscript{a,b,}\thanks{axel.molle@kuleuven.be} , 
	Oleg Zatsarinny\textsuperscript{c,}\thanks{author deceased},
	Thomas Jagau\textsuperscript{%
		{d}}, 
	Alain Dubois\textsuperscript{a} and 
	Nicolas Sisourat\textsuperscript{a,}\thanks{nicolas.sisourat@upmc.fr}
	\\
	\begin{minipage}[t]{\textwidth}\medskip\small
	\begin{description}%
	\item[\textsuperscript{a}] Laboratoire de Chimie Physique - Mati\`ere et Rayonnement, Sorbonne Universit\'e / CNRS, F-75005 Paris, France
	\item[\textsuperscript{b}] %
	{Institute for Theoretical Physics, Department of Physics and Astronomy, KU Leuven, B-3001 Leuven, Belgium}
	\item[\textsuperscript{c}] Department of Physics and Astronomy, Drake University, Des Moines, IA 50311, USA
	\item[%
	{\textsuperscript{d}}] %
	{Quantum Chemistry and Physical Chemistry, Department of Chemistry, KU Leuven, B-3001 Leuven, Belgium}
\end{description}
\end{minipage}
}
\date{13\textsuperscript{th} December 2022, revised 1st Feb 2023}

\providecommand{\citeleiden}{\autocite{heays-aa2017, vanharrevelt-jpca2008, mordaunt-jcp1994, fillion-jphysb2003, fillion-jcp2004, mota-chemphyslett2005, smith-apj1981, chan-chemphys1993, yoshino-chemphys1996, yoshino-chemphys1997, parkinson-chemphys2003, fillion-jphysb2003, fillion-jcp2004}}
\begin{document}

\maketitle

\begin{abstract}
A decade ago, an electron-attachment process called \textit{interatomic coulombic electron capture} has been predicted to be possible through energy transfer to a nearby neighbour. It has been estimated to be competitive with environment-independent photorecombination but its general relevance has yet to be established.
Here, we evaluate the capability of alkali and alkaline earth metal cations to capture a free electron by assistance from a nearby water molecule.
We introduce a characteristic distance $\dist_\IC$ for this
energy transfer mechanism in equivalence to the F\"orster radius. Our results show that
 water-assisted electron capture
dominates over photorecombination beyond the second hydration shell of each
cation for electron energies above a threshold.
{
	The assisted capture reaches distances equivalent to a 5th to 7th solvation shell for the studied cations.
}
The far reach of
assisted electron capture is of significant general interest to the broad spectrum of research fields
dealing with low-energy electrons, in particular radiation-induced damage of biomolecules. 
The here introduced distance measure will enable to quantify the role of the environment for assisted electron attachment.
\end{abstract}

\newcommand{\watthre}{12.54~eV}
\newcommand{\watemax}{13.41~eV}
\newcommand{\watdmax}{1.45306~nm}
\newcommand{\Lithre} {7.16~eV}
\newcommand{\Lidmax} {1.39~nm}
\newcommand{\Liemax}{12.45~eV}
\newcommand{\LirI}{0.19~nm}
\newcommand{\LirII}{0.43~nm}
\newcommand{\Lircite}{\autocite{kiyohara-jcp2019,loeffler-jcp2002,lyubartsev-jcp2001,howell-jpcm1996,egorov-jml2000}}
\newcommand{\Bethre}{-5.64~eV}
\newcommand{\Bedmax} {1.25~nm}
\newcommand{\Beemax}{12.02~eV}
\newcommand{\BerI}{0.17~nm}
\newcommand{\BerII}{0.37~nm}
\newcommand{\BerIII}{0.54~nm}
\newcommand{\Bercite}{\autocite{smirnov-rjgc2008,rudolph-dalton2009}}
\newcommand{\Nathre} {7.50~eV}
\newcommand{\Nadmax} {1.41~nm}
\newcommand{\Naemax}{12.52~eV}
\newcommand{\NarI}{0.24~nm}
\newcommand{\NarII}{0.44~nm}
\newcommand{\Narcite}{\autocite{galib-jcp2017,kiyohara-jcp2019}}
\newcommand{\Mgthre}{-2.32~eV}
\newcommand{\Mgdmax} {1.33~nm}
\newcommand{\Mgemax}{12.14~eV}
\newcommand{\MgrI}{0.20~nm}
\newcommand{\MgrII}{0.41~nm}
\newcommand{\Mgrcite}{\autocite{kiyohara-jcp2019}}
\newcommand{\Kthre} {8.20~eV}
\newcommand{\Kdmax} {1.43~nm}
\newcommand{\Kemax}{12.24~eV}
\newcommand{\KfourS}{14.22~eV}
\newcommand{\KfourP} {9.88~eV}
\newcommand{\KrI}{0.26~nm}
\newcommand{\KrII}{0.49~nm}
\newcommand{\Krcite}{\autocite{kiyohara-jcp2019}}
\newcommand{\Cathre} {0.82~eV}
\newcommand{\Cadmax} {1.40~nm}
\newcommand{\Caemax} {5.56~eV}
\newcommand{\CathreeD}{3.13~eV}
\newcommand{\CarI}{0.23~nm}
\newcommand{\CarII}{0.45~nm}
\newcommand{\Carcite}{\autocite{kiyohara-jcp2019}}

\newcommand{\LiSons}{1.15~nm}
\newcommand{\LiPons}{1.32~nm}
\newcommand{\LiPthr}{9.00~eV}
\newcommand{\Lidend}{1.23~nm}
\newcommand{\NaSons}{1.06~nm}
\newcommand{\NaPons}{1.27~nm}
\newcommand{\NaPthr}{9.54~eV}
\newcommand{\Nadend}{1.28~nm}
\newcommand{\KSons}{0.90~nm}
\newcommand{\KSthr}{8.33~eV}
\newcommand{\KPons}{1.28~nm}
\newcommand{\KPthr}{9.82~eV}
\newcommand{\Kdend}{1.26~nm}

\section{Introduction}
\label{s:intro}

\paragraph{}
Water is the vital prerequisite for life on Earth. Understanding its interaction with the respective solute is therefore essential. An important class of solutes is minerals that  are dissolved in their ionic form and can be classified thereby. The alkali metals lithium, sodium and potassium are singly positively charged in their dissolved form in water, the alkaline earth metals beryllium, magnesium and calcium are doubly charged. 

The right oxidation number is important for many chemical reactions within biological organisms.  Changing the oxidation state of an ion to a biochemically more advantageous form is not straightforward for the organism. Radiation experienced for instance from the sun, xrays or radioactive material can change directly or indirectly the oxidation state of irradiated elements.

The \textit{interatomic coulombic electron capture} (ICEC) effect is a 
less-investigated example among the various processes that can lead to a change of oxidation number.
Contrary to resonant electron thermalisation, ICEC works in support of an electron attachment by assistance of surrounding atoms and molecules. 
We show in this work that the mere presence of a solvent water molecule can make a significant contribution to an increased attachment probability of slow electrons to dissolved nutrient ions. This can have a considerable effect on bioavailability of the nutrients as well as on the propagation of free charges through the organism in the wake of initial ionising irradiation.
\begin{figure}[h!tb]\centering
	\includegraphics[]{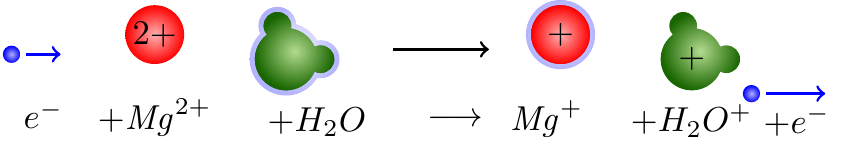}
	\caption{Schematic reaction of interatomic coulombic \textcolor{blue!50!black}{{electron}} capture (ICEC) by a magnesium~(II) \textcolor{red!60!black}{{cation}} $\mathit{Mg}^{2+}$ through ionisation of a nearby \textcolor{green!40!black}{{water molecule}} $H_2O$. 
	The \textcolor{red!50!black}{{cation}} recombines with a free \textcolor{blue!50!black}{{electron}} $e^-$ to form a singly charged \textcolor{red!50!black}{\textbf{ion}}. The excess energy is transferred to the \textcolor{green!40!black}{{water molecule}} which is ionised. 
	This leads to a \textcolor{green!40!black}{{water molecule}} cation and a free \textcolor{blue!50!black}{{electron}} emitted from the water molecule with \textcolor{blue!50!black}{{different velocity}}.}
	\label{f:ICEC}
\end{figure}

\paragraph{}
Interatomic coulombic electron capture is a non-local energy transfer process facilitating recombination of a free electron with an ion by ionisation of a neighbour. Schematically depicted in \autoref{f:ICEC}, an electron can attach for instance to a magnesium~(II) cation by transfer of {excess energy} to a nearby water molecule. 
Water then releases another electron in order to rid itself from the energy.
For these reaction partners, this process leaves both species positively charged and emits the propagating electron faster than the initial one. 
In context of a dissolved alkali or alkaline earth metal $A$ which appears in its ionic form of charge $(q+1)$ in water, interatomic coulombic electron capture can generally be expressed as
\begin{equation}
e^- +A^{+1(+q)} \;\;+H_2O \longrightarrow A^{0(+q)} \;\;+H_2O^+ +e^-
\end{equation}
where $A$ can be lithium $\mathit{Li}$, sodium $\mathit{Na}$ or potassium $\mathit{K}$ for alkali metals with $q=0$, and beryllium $\mathit{Be}$, magnesium $\mathit{Mg}$ or calcium $\mathit{Ca}$ for alkaline earth metals with q=1. Particularly since it is aided by a molecule of the solvent agent water here, one may call the process similarly \textit{environment-assisted electron capture}.{\autocite{gokhberg-jphysb2009}}

ICEC is emerging as a research field.\autocite{icdrefbase} However, experimental investigations are yet lacking. Furthermore, the computation of ICEC observables is a challenge. The virtual-photon approximation is a robust asymptotic formula, that allowed the initial postulation of the existence of ICEC in the year 2009.\autocite{gokhberg-jphysb2009,gokhberg-physrev2010}
So far, this is the only approach that is technically able to handle systems relevant to biology.
The molecular R-matrix approach is being explored and has been successful for very small molecular systems.\autocite{sisourat-pra2018} It has shown that the virtual-photon approximation provides a lower limit and provides the correct trend but omits overlap and interference of wavefunctions as well as molecular distortions from intermolecular close-range interactions.\autocite{molle-pra2021,molle-pra2021b}

Beyond the molecular aspect, electronic dynamics of ICEC have been investigated successfully in relation to nanowires and embedded quantum dots.\autocite{pont-physrev-2013,bande2013-04031,pont-jpcm-2019,molle-jcp-2019, molle-phd-2019} In a mean-field approach, it has been proposed possible in macroscopic trapped cold-atom systems.\autocite{molle-phd-2019}
Thereby manifesting a fundamental process in various fields of interest, the original naming of the process as \enquote{interatomic coulombic electron capture} has seen its expansion to include \enquote{intermolecular}, \enquote{inter-quantum dot}, or in an attempt to generalize the term inclusively to any two subsystems, to \enquote{interparticle coulombic electron capture} under the same acronym of ICEC. 

In this work, we investigate the ICEC process in microhydrated cations using the virtual-photon approximation. Based on this approach, our results show that the presence of water molecules increases significantly the electron attachment cross sections to the cations due to ICEC. Furthermore, we introduce a characteristic distance $\dist_\IC$ for ICEC and demonstrate that the latter dominates over photorecombination beyond the second hydration shell of each cation for electron energies above a threshold.

The paper is organized as follows: In \autoref{s:theo}, we present the theoretical approach employed in this work. The computational details are provided in \autoref{s:computation}. In \autoref{s:results}, we discuss the results for alkali monocations and alkaline earth dications. Finally, the conclusions of this work are reported in \autoref{s:conclusion}.

\section{{Theoretical Derivation}}\label{s:theo}
\
\label{s:theory}

Intermolecular coulombic electron capture has been first investigated in the virtual photon approxi\-mation~\autocite{gokhberg-jphysb2009,gokhberg-physrev2010}. %
{Within this approach and for the systems considered here, the corresponding cross section for  the electron capture into a specific state $i$ of the metal ion is given by
\begin{equation}
\sigma_{\IC_i} = \frac{3}{4\pi \dist^6} \, \left(\frac{\hbar c}{h\nu}\right)^{4} {\sigma_\wat} \, \sigma_{\rec} . \label{eq:s_IC_i}
\end{equation}
}
In the above equation, $\sigma_{\rec}$ is the partial photorecombination cross section of $A^+$ and $\sigma_\wat$ is the water photoionisation cross section. The exchanged energy $h\nu$ is the sum of the free electron energy $\epsilon$ and the ionisation potential $V_{\ato}$ of the capturing state, $h\nu =\epsilon +V_{\ato}$, due to energy conservation. The distance between the two partners is noted $r$. The total ICEC cross section is thus
\begin{equation}
\sigma_{\IC} = \sum_i \sigma_{\IC_i}
\label{eq:s_IC}
\end{equation}
where the sum runs over %
{all open ICEC channels}.

Note that these equations accommodate the possibility of local resonances in the molecular subsystems described by the respective cross section $\sigma_{\wat}$ for the assisting partner, and $\sigma_{\rec}$ for the recombining cation. 
{However, the interactions between the partners are neglected.}
In its assumption of distinguishable subsystems, it can be characterised as an asymptotic formula.

Interpreting the partial cross section for assisted capture $\sigma_{\IC_i}$ as function of the photorecombination cross section $\sigma_{\rec}$, their ratio $(\sigma_{\IC_i} / \sigma_{\rec})$ expresses the amplification factor arising from the assisting effect of the mitigating partner, in this case the water molecule. This ratio is by its nature, a numerical coefficient without physical unit.

By rearrangement of the quantities in Eq.~\eqref{eq:s_IC_i},  
\begin{equation}
	\left(\frac{\sigma_{\IC_i}}{\sigma_{\rec}}\right)\, \dist^6  = \frac{3}{4\pi } \, \left(\frac{\hbar c}{h\nu}\right)^{4} {\sigma_\wat} \mathrel{\hat=} (distance)^6,
	\label{eq:dist_i^6}
\end{equation}
we therefore identify a quantity representing a length-scale purely on the argumentative grounds of consistent physical units. Implicitly, the transferred energy $h\nu$ and consequently the argument to the photoionisation cross-section $\sigma_{\wat}(h\nu)$ both depend on the energy released by the specific capturing state $i$. This fact shall be indicated in the following by explicitly stating the index $i$ on those quantities.

We can interpret the identified distance of Eq.~\eqref{eq:dist_i^6} as a characteristic length
\begin{equation}
	\dist_{\IC_i} := \left( \frac{3}{4\pi } \, \left(\frac{\hbar c}{h\nu_{i}}\right)^{4} {\sigma_\wat^{(i)}} \right)^{\frac16}
	\label{eq:r_IC_i}
\end{equation} 
 for water-assisted electron capture into the capturing state $i$. 
 The ratio of this parameter $\dist_{\IC_i}$ over a particular distance $r$ between the recombining partner and the assisting water molecule, $(\dist_{\IC_i} / \dist)$, is then equivalent to the amplification factor $(\sigma_{\IC_i} / \sigma_{\rec})$ in terms of the respective cross sections as
 \begin{equation}
 	\frac{ \sigma_{\IC_i} }{ \sigma_{\rec} }
 	= \left(\frac{\dist_{\IC_i} }{ \dist }\right)^6 \;.
 \label{eq:s_i/s_i}
 \end{equation}
 In this sense, the present water molecule can be seen as \emph{stimulating} the recombination of cation and free electron. The specific characteristic length $\dist_{\IC_i}$ introduced here is thereby indicating the intermolecular distance
 at which the partial cross section for assisted capture into a specific capturing state $i$ is of equal magnitude to the partial cross section of photorecombination into the same capture state. Note that this characteristic length $\dist_{\IC_i}$ does not depend on the photorecombination cross section itself.
 
 The additivity in the individual cross sections allows to similarly define a total characteristic distance $\dist_{\IC}$ for water-assisted capture.

Introducing the partial photorecombination cross sections as statistical weights
\begin{equation}
w_i := \frac{ \sigma_{\rec} }{ \left( \sum_j \sigma_{\rec[j]} \right) } \; ,
\label{eq:w_i}
\end{equation}
such that each $w_i \leq 1$ for any $i$ and their sum
$\sum_i w_i \equiv 1$ for any electron energy, we find an
 
overall total characteristic length
\begin{equation}
	\dist_{\IC} := \left( \sum_i w_i \,(\dist_{\IC_i})^6 \right)^{\frac16}.
	\label{eq:r_IC(r_IC_i)}
\end{equation}

As a consequence, the competitive impact of the overall water-assisted electron capture with respect to the environment independent photo-recombination can thus be expressed 
by the ratio
\begin{equation}
\frac{\sigma_{\IC}}{\sigma_{\PR}}
= \left( \frac{\dist_{\IC}}{\dist}\right)^6
\end{equation}
between the total ICEC cross section and the total photorecombination cross section. If the capturing cation is closer to the water molecule than the distance $\dist_\IC$ which is a function of incident energy $\epsilon$, then ICEC dominates the environment-independent photorecombination.

The quantum yield, or efficiency of the environment-assisted electron capture with respect to the total electron capture from both processes is thus distance-dependent as
\begin{equation}
	\eta_\IC = \frac{\sigma_{\IC}}{\sigma_{\PR} + \sigma_{\IC}} = \frac{\dist_\IC^6}{\dist^6 + \dist_\IC^6} \;.
\end{equation}

The characteristic distance $\dist_\IC$ can be interpreted analogously to the F\"orster radius in the case of intermolecular energy transfer between two fluorescent molecules. Known as \textit{F\"orster resonant energy transfer} (FRET) for bound electronic excitations, the same distance dependence with respect to a characteristic length arises there.\autocite{foerster-andp1948} 
The characteristic distance $\dist_\IC$ is thereby exactly that distance between one electron captor and its reaction partner at which the efficiency $\eta_\IC$ of ICEC measures 50\%. This means the partner-assisted capture cross section at this distance has equal magnitude to that of the environment-independent photorecombination.

{
    Note that this is a definition for a single reaction partner. In an environment with multiple independent partner molecules counted with index $N$, each will contribute to the overall many-partner ICEC cross section in the environment, $\sum_N \sigma_{\IC}\left( \dist_N \right)$, in dependence on its individual distance $\dist_N$ from the energy donor. The introduced characteristic distance $\dist_{\IC}$ remains a pairwise measure independent of the individual partner distance. In the following, we consider the cases of only one partner molecule. The cross sections and ICEC radius reported represent therefore lower limits.
}

While photorecombination data is sometimes hard to come by, photoionisation cross sections are often well tabulated.
Let $g_\cat$ denote therefore the statistical weight of the alkali cation describing the number of electronic states equivalent to the initial state, and let $g_{\ato}$ be the equivalent multiplicity of the $i$th capturing state, then the incident electron energy and emitted photon energy relate the photorecombination cross section as
{to the photoionisation cross section $\sigma_{\ion}$ through the \textit{principle of detailed balancing},\autocite{fowler-pnas1925},}
\begin{equation}
(2m_e{c^2}) \,\epsilon \, g_{\cat} \, \sigma\!_{\rec} = {(h\nu_i)^2} \, g_{\ato} \, \sigma\!_{\ion} \;,
\end{equation}
known in this case as the Milne relation.\autocite{oxenius1986} %
{This relation has also been used by \citeauthor{gokhberg-jphysb2009} %
to reformulate the partial cross section for assisted electron-capture in terms of the more accessible photoionisation cross section of (excited) state $i$.\autocite{gokhberg-jphysb2009}}

The discussed quantities represent functions of the energy, either directly through the continuum energy $\epsilon$ of the captured electron and the ionisation threshold $V_{\ato}$ associated with the capturing state $i$, or indirectly through their energy difference transferred as photon energy $h\nu$.

Here, $V_{\ato}$ represents a discrete set of energies and $\epsilon$ a particular value within the energy continuum. Numerically, however, $\epsilon$ is usually represented by a finite discrete collection of values together with partial photo{recombination} cross sections at that value $\sigma_{\rec}(\epsilon)$%
{, or as associated photoionisation cross sections $\sigma_{\ion}(\epsilon+V_{\ato})$}. In the following, we therefore indicate the composite index $(\iep)$ to remind the reader of the implicit energy dependence on both incident-electron energy $\epsilon$ as well as ionisation threshold $V_{\ato}$.
Taking care of the interdependence of each photoionisation cross section through the exchanged energy $h\nu_\iep$ 
the total assisted capture cross section itself may be estimated as
a weighted sum
\begin{equation}
	{\dist_{\IC_\epsilon}}
		= \left( \frac{ 3\,({\hbar c})^4}{4\pi}  \,
		\sum_i w_{\iep} \,
		\frac{\sigma_{\wat}^{(\iep)}}{(h\nu_\iep)^{4}}
		\right)^{\frac16}
		\,
	\label{eq:d_IC}
\end{equation}
{where the statistical weights in terms of partial photoionisation cross sections take the form
\begin{equation}	
w_{\iep} = 
	\frac{
		{(h\nu_\iep)^2} \, {g_{\ato[i]}} \; \sigma\!_{\ion[\iep]}
	}{
		\sum_j {(h\nu_\jep)^2} \, {g_{\ato[j]}} \; \sigma\!_{\ion[\jep]}
	}\label{eq:w_iep} \;.
\end{equation}
These weigths are in themselves independent of the reaction partner and mix the available capturing states to form the sum $\sum_i w_{\iep} \equiv 1$ for any incident electron energy $\epsilon$ while the fraction of ionisation cross section ${\sigma_{\wat}^{(\iep)}}$ of the reaction partner over the fourth power of transferred energies $h\nu_{\iep}$ determine the scale of characteristic distance $\dist_\IC$. Although Eq.~\eqref{eq:w_i} appears simpler, it represents the same quantity as Eq.~\eqref{eq:w_iep} linked through the Milne relation and the latter has been used in conjunction with the available photoionisation data to compute $\dist_{\IC}$.
\

\subsection{Relevant Quantitative Limits}
In the following, we examine the upper bound in magnitude, as well as the near-threshold and the large-energy behaviour of the introduced characteristic distance $\dist_{\IC}$.
These findings are of value to estimate more generally the viability of a potential experimental investigation: What length scale is to be expected, how does it behave for very low electron energies, what is to be expected for high electron energies?
These questions arise immediately when evaluating whether a certain experimental setup may allow to measure ICEC. 

\begin{enumerate}
	\item \textbf{Upper bound:} The specific characteristic distance $\dist_{\IC_i}$ for assisted capture into state $i$ as given by Eq.~\eqref{eq:r_IC_i} is independent of the electron-capturing species, i.e. it solely depends on the photoionisation cross section $\sigma_{\wat}$ of the assisting water molecule as function of photon energy $h\nu$.
	In general, this photoionisation cross section is a finite positive quantity. 
	It vanishes for energies below the ionisation threshold.
	For the water molecule, this ionisation threshold is 12.6~eV.\autocite{potts-prsl1972} 
	Similarly, the photoionisation cross section also tends to approach zero with increasing energy in the high-energy regime. 
	This implies, there exists a global maximum.
	More particularly, we are interested in the global maximum of the (auxiliary) function
	\begin{equation}
		f(h\nu) := \frac{ \sigma_{\wat}(h\nu) }{ (h\nu)^4 } \,, \text{ since } \dist_{\IC_i}(h\nu) = \left( \frac{3 (\hbar c)^4}{4\pi} \,f(h\nu) \right)^{\frac16} \,.
		\label{eq:f(hv)}
	\end{equation}
	We can therefore define the length
	\begin{equation}
		\dist_{\max} := \left( \frac{3 (\hbar c)^4}{4\pi} \, \max_{h\nu}\left[f(h\nu)\right] \right)^{\frac16}
		\label{eq:r_max}
	\end{equation}
	through the global upper bound of the function $f$.
	{This upper bound is by its definition independent of the kinetic energy of the electron incident on the cation.}
	Independent of the particular index $i$ for the electron-capturing state, any specific characteristic distance $\dist_{\IC_i}$ is thereby bound from above, as
	\begin{equation}
		\dist_{\IC_i}(h\nu) \leq \dist_{\max}
		\,.
	\end{equation}
	This has a direct implication on the total characteristic distance which represents according to Eq.\eqref{eq:r_IC(r_IC_i)} a weighted sum over the bound quantities $\left\{\dist_{\IC_i}\right\}$ as 
	\begin{equation}
		\left(\dist_{\IC}\right)^6 = \sum_i w_i \left(\dist_{\IC_i}\right)^6 
		\leq \left(\dist_{\max}\right)^6 \sum_i w_i \,.
	\end{equation}
	By their definition, the sum over weights $\left\{w_i\right\}$ is unity at any (photon) energy where at least one electron-capture channel is open,
	\begin{equation}
		\sum_i w_i =
		\begin{cases}
			1 \text{, if } \exists i\in \mathbb{N} \mid w_i \neq 0\\
			0 \text{, if } w_i = 0 \mathrel \forall i
		\end{cases}.
	\end{equation}
	Hence, the total characteristic distance $\dist_{\IC}$ is bound from above,
	\begin{equation}
		\dist_{\IC} {(\epsilon)} \leq \dist_{\max} \,,
	\end{equation}
	for all energies by the same quantity $\dist_{\max}$ as each individual specific distances $\dist_{\IC_i}$ associated with electron-capture state $i$.
	{Note that $\dist_{\max}$ is independent of the particular electron-captor.} This suggests $\dist_{\max}$ as an easily accessible quantity to estimate the length scale associated to assisted electron-capture for any given assisting partner species. The bound on the length-scale of assisted capture is defined solely by the stimulating partner.
	
	\item \textbf{Low energies:}
	In the context of energy exchange through assisted-electron capture, the photoionisation cross sections of electron captor and assisting partner are linked through the exchanged photon energy $h\nu$. This quantity is however dependent on the kinetic energy of the incident electron $\epsilon$ which is a continuous degree of freedom, as well as the ionisation potential $V_{\ato}$ of the particular capturing state $i$ which is a discrete degree of freedom, 
	\begin{equation}
		h\nu = \epsilon + V_{\ato} \,.
	\end{equation} 
	We do not have a direct control %
	which state captures the incident electron but rather have to be aware that in the general case, there is a set of capturing states $\left\{i\right\}$ with a discrete set of transferred energies $\left\{h\nu_{\iep}\right\}$ for any fixed incident energy $\epsilon$.
	In order to meet the ionisation threshold $V_\wat$ of the assisting partner, the kinetic energy of the free electron needs to fulfil the criterion
	\begin{equation}
		\epsilon = h\nu - V_{\ato} \geq V_{\wat} - V_{\ato}
	\end{equation}
	for at least one capture state.
	The sign of the difference in ionisation potentials indicates whether the energy transfer is endo- or exothermic, in other words, whether the energy accepting electron on the assisting partner is emitted with a lower, or respectively higher kinetic energy than $\epsilon$.
	We assume without loss of generality, that the capturing state index $i$ is ordered by the respective ionisation potential with $V_{\ato[0]}\leq V_{\ato[1]}\leq ...$ .
	Then $i=0$ marks an electron-capture into the ground state and the energy difference 
	\begin{equation}
		V_{\wat} - V_{\ato[0]} =: \epsilon_0
	\end{equation}
	represents the energy threshold for assisted electron capture.
	Depending on the choice of reaction partners, this quantity can be positive, i.e. $\epsilon_0 > 0$. Then the free electron needs a kinetic energy of at least $\epsilon_0$ to allow for assisted capture to occur. This is the case for the alkali cations $Li^+$, $Na^+$, and $K^+$ where $\epsilon_0$ ranges from 7.16 eV\autocite{mendoza-physcr1996,NIST_ASD,bushaw-pra2007} to 8.20 eV\autocite{NIST_ASD,LORENZEN1981370, C-JLorenzen_1983,7185EL,zatsarinny-physreva2010,yar-physreva2013} with respect to water.\autocite{potts-prsl1972} \citeleiden In the case where $\epsilon_0$ vanishes, the energy transfer between captor and assisting energy acceptor is energy neutral. Similarly, the energy threshold can be negative, i.e. $\epsilon_0<0$. This is the case for electron-capture by alkaline earth cations $Be^{2+}$ and $Mg^{2+}$ in assistance from water.\autocite{NIST_ASD,AEKramida_2005,doi:10.1063/1.555879,mendoza-physcr1996} In that case, even an infinitely slow free electron, i.e. $\epsilon=0$, allows for an assisted capture into at least one capturing state.
	Moreover, a negative threshold suggests that even pseudo-free electrons, for example from Rydberg states with ionisation potentials smaller than $|\epsilon_0|$, may energetically be captured through the mechanism of assisted electron capture. In that respect, ICEC merges over to a related process known as electron-transfer-mediated decay (ETMD(3)).\autocite{Zobeley01} 
	
	At an advantageous choice of both reaction partners, the characteristic distance may already have a significant size at vanishing electron energy, %
	{ $\epsilon = 0$. That means there is in that case at least one capture state $i$ for which $w_\iep>0$.} 
	Each state-specific characteristic distance is governed by the same function $f(h\nu)$ of transferred photon energy $h\nu$. In the limit of an infinitely-slow incident electron, where $\epsilon=0$, the transferred energy $h\nu=\epsilon+V_{\ato} = V_{\ato}$, reduces to the specific ionisation potential of the respective capturing state $i$. The characteristic distance for assisted capture is purely determined by the water %
	{molecule}'s photoionisation cross section at the energy of the specific ionisation potential as
	\begin{equation}
		\dist_{\IC_i} (\epsilon=0) = \left( \frac{ 3 (\hbar c)^4 }{ 4\pi } \,f(V_{\ato}) \right)^{\frac16} \,.
	\end{equation}

Since $f$ is positive and bound from above, there is a particular capture state $k$ which represents the biggest value in the discrete set of function values $\{f(V_{\ato})\}$, such that
	\begin{equation}
		f(V_{\ato}) \leq \sup_i \left\{ f( V_{\ato} ) \right\} = f(V_{\ato[k]}) \text{ for a particular } k \in \{i\}.
	\end{equation}
	The specific characteristic distance for every state is therefore limited by that distance $\dist_{\sup} =: \dist_{\IC_k} (\epsilon=0)$ of the identified capture state $k$ such that for any state $i$
	\begin{equation}
		\dist_{\IC_i} (\epsilon=0) \leq \dist_{\sup} \leq \dist_{\max} \,.
	\end{equation}
	The total characteristic distance for assisted electron-capture being a weighted sum over the individual capturing states is thereby also bound. Since
	\begin{equation}
	\left(\dist_{\IC} (\epsilon=0)\right)^6 
	= 
	\sum_i w_{i} \,(\dist_{\IC_i})^6
	\leq 
	(\dist_{\sup})^6 \,\sum_i w_{i} \,,
	\end{equation}
	the total characteristic distance is bound by the single biggest contribution $\dist_{\sup}$ in the discrete set $\{\dist_{\IC_i}\}_i$, so that also
	\begin{equation}
		\dist_{\IC} (\epsilon=0) 
		\leq \dist_{\sup} \leq \dist_{\max} \,.	
	\end{equation}

	\item \textbf{High energies:}
In the context of this work, the regime of high energies is reached if the incident electron energy $\epsilon$ is much greater than the largest ionisation threshold $V_{\ato[0]}$ in the energy-ordered set of thresholds $\{V_{\ato}\}$ respective to the discrete quasi-infinite set of capture states $i$. The exchanged energy between electron-captor and assisting partner molecule, 
	\begin{equation}
		h\nu_{0} ({\epsilon\gg V_{\ato[0]}}) = \epsilon + V_{\ato[0]} \approx \epsilon + V_{\ato[1]} \approx \ldots \approx \epsilon + V_{\ato} + \ldots \approx \epsilon
	\end{equation}
	is therefore approximately equal to the pure kinetic energy of the incident electron. 

	This implies that in the high energy limit, every specific characteristic distance $\dist_{\IC_i}$ asymptotically approaches the same function in energy $\dist_{\infty}(\epsilon)$.
	The overall asymptotic limit is thereby also given by 
	\begin{equation}
		\lim_{\{V_{\ato}\} \ll \epsilon} \dist_{\IC} (\epsilon)
		= \dist_{\infty} (\epsilon) %
		=\left( \frac{3(\hbar c)^4}{4\pi} \,f(\epsilon)\right)^{\frac16} \leq \dist_{\max}
		\,.
		\label{eq:r_infty}
	\end{equation}
	That implies that the total characteristic distance for assisted electron-capture becomes independent of the specific captor in the high energy regime.
\end{enumerate}

\section{{Methods}}\label{s:computation}
\

We have employed the virtual-photon approximation to compute the ICEC cross sections.\autocite{gokhberg-jphysb2009,gokhberg-physrev2010} 
{
    This method assumes distinctly separated subsystems. In consistence with this approximation, all quantities used for cations and water molecules are therefore for the systems in the gas phase (i.e. isolated).
} 

The evaluation of the characteristic distance and total cross section
requires the set of photoionisation data for ground and excited states of the captor according to Eq.~\eqref{eq:d_IC}. 
{%
The results for the characteristic distance are therefore to be seen as an asymptotic result. We stress again that they include possible local resonances as far as they are covered by the respective photoionisation cross section of the individual reaction partner.
}

Below we report the databases employed in this work and the procedure followed to interpolate and extrapolate the missing data.

\subsection{Partial Photoionisation Cross Sections of Metals}
To our current knowledge, the most extended consistent set of excited state ionisation cross sections of atoms and ions is provided by the \textsc{top}base\autocite{mendoza-physcr1996} dataset which is a purely theoretical database of R-matrix calculations.
It allows us to gather data for 25 capturing states of the lithium cation $Li^+$, 33 capturing states for the sodium cation $Na^+$, 25 capturing states for the beryllium cation $Be^{2+}$, 33 capturing states for the magnesium cation, $Mg^{2+}$ and 36 states for the calcium cation $Ca^{2+}$. Potassium is not available within this database. 

Computations with the Dirac-based B-spline R-matrix\autocite{zatsarinny-cpc2006,zatsarinny-physreva2008} have been tried in 2010 outside the \textsc{top}base project,\autocite{zatsarinny-physreva2010} and were experimentally confirmed later.\autocite{yar-physreva2013} This allows to use consistent data of 14 capturing states for the potassium cation $K^+$ which would otherwise not be possible.

\subsection{Photoionisation Cross Sections of Water}
\providecommand{\citeleiden}{\autocite{heays-aa2017, vanharrevelt-jpca2008, mordaunt-jcp1994, fillion-jphysb2003, fillion-jcp2004, mota-chemphyslett2005, smith-apj1981, chan-chemphys1993, yoshino-chemphys1996, yoshino-chemphys1997, parkinson-chemphys2003, fillion-jphysb2003, fillion-jcp2004}}
For the assisting water molecule, photoionisation cross section data for the ground-state ionisation is sufficient. It is advantageous if the photoionisation cross section is provided as function of photon energy $h\nu$ instead of emitted electron energy. 
This allows to treat the ionisation of the water molecule as a single function in accord with Eq.~\eqref{eq:d_IC}. 
The Leiden database for photodissociation and photoionisation of astrophysically relevant molecules has been used.\autocite{heays-aa2017} Within this database, data for the water molecule stem from a considerable number of experimental sources.\citeleiden
This dataset has been used to arrive at an estimate for the order of the characteristic distance of the ICEC process assisted by the water molecule according to Eqs.~\eqref{eq:d_IC} \& \eqref{eq:r_max}. Therefore, we expect a characteristic distance of the order
\begin{multline}
	\mathcal{O}(\dist_\IC) \leq 
		\max\limits_\epsilon
		\left[
			\left( 
				\frac{3\,({\hbar c})^{4}}{4\pi}  \,
				\frac{\sigma_{\wat_{\epsilon}}}{\epsilon^4}
			\right)^{\hskip-1ex\frac16}
		\right] \\= 1.0091656 \,\frac{\text{nm} ~\text{Ryd}^{\frac46}}{\;\text{Mb}^{\frac16}} 
		~\max\limits_\epsilon\left( \frac{\sigma_{\wat_{\epsilon}}}{\epsilon^4} \right)^{\hskip-.1ex\frac16}\;.
\end{multline} 
which shows that a natural choice of units for the characteristic distance is nanometres when energies are handled in Rydberg and cross sections in megabarn. These are thus the employed units for the calculations even though we present energies in electron volts throughout the discussion. Note that this estimate is independent of the electron captor itself. As consequence of the available data set, the magnitude of characteristic distance $\dist_\IC$ for water-assisted ICEC are maximally $r_{\max}=1.45306~\mathrm{nm}$ which corresponds to a photon energy of about \watemax.\citeleiden

\subsection{Interpolation and Extrapolation}
The tables for various capturing states may provide data points at differing incident electron energies which lead to mismatching photon energies in the data table for the water molecule. Necessarily, interpolation for intermediate energies between the given data points is necessary. 
We have interpolated linearly to intermediate points. Where necessary data was missing at the larger energy range, a capturing state's photoionisation cross section has been extrapolated from the last 10\% of available data but at least the 10 last data points. For simplicity and in accordance with a high-energy power law of proportionality to $\epsilon^{-3.5-\ell}$,\autocite{verner-apj1996} we extrapolated unavailable data by a simple power law as $\ln \sigma = a + b \ln \epsilon$ with extrapolation parameters $a$ and $b$. 

\section{Results and Discussion}\label{s:results}

\begin{table*}\centering
	\caption{Key values of the distance $\dist_{\IC}$ for water-assisted electron capture.}\label{t:rIC}
	\begin{tabular}{r r r c l}
		\toprule
		species & threshold energy & optimal energy& maximal radius & first hydration shell \\
		$\cat$ & $V_{\ato[]} - V_\wat$ [eV] & $\epsilon$ where $\dist_{\IC} = \max\!.$ [eV] & $\max\limits_\epsilon \dist_{\IC}$ [nm]& $r_1$ [nm]\\
		\midrule
		$Li^+$ & \Lithre & \Liemax & \Lidmax & \LirI\Lircite\\
		$Na^+$ & \Nathre & \Naemax & \Nadmax & \NarI\Narcite\\
		$K^+$ & \Kthre & \Kemax & \Kdmax& \KrI\Krcite\\
		$Be^{2+}$ & \Bethre & \Beemax & \Bedmax & \BerI\Bercite\\
		$Mg^{2+}$ & \Mgthre & \Mgemax & \Mgdmax& \MgrI\Mgrcite\\
		$Ca^{2+}$ & \Cathre & \Caemax & \Cadmax & \CarI\Carcite\\
		\bottomrule
	\end{tabular}
\end{table*}

\begin{figure*}[h!tb]
	\centering
	{****}
	\includegraphics[page=1,width=0.5\textwidth]{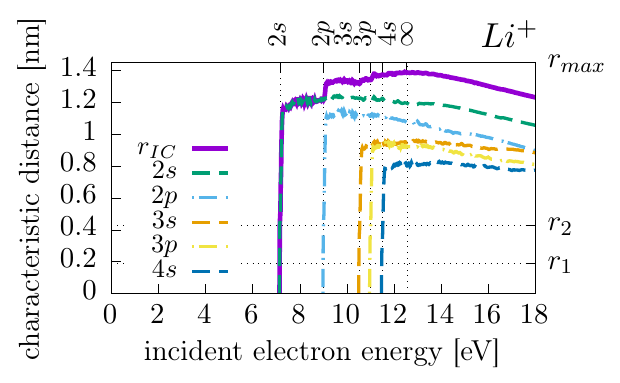}%
	\hspace*{-.5em}\hfill%
	\includegraphics[page=2,width=0.5\textwidth]{figure2.pdf}
	
	\includegraphics[page=3,width=0.5\textwidth]{figure2.pdf}%
	\hspace*{-.5em}\hfill%
	\includegraphics[page=4,width=0.5\textwidth]{figure2.pdf}
	
	\includegraphics[page=5,width=0.5\textwidth]{figure2.pdf}%
	\hspace*{-.5em}\hfill%
	\includegraphics[page=6,width=0.5\textwidth]{figure2.pdf}
	{****}
	
		\caption[$\dist_{\IC}$ for electron capture by alkali and alkaline earth cations assisted by {\wat}]
		{
			Characteristic distance $\dist_{\IC}$ of ICEC by %
			alkali and alkaline earth cations in assistance from a water molecule as function of incident electron energy $\epsilon$. Key numbers of energy threshold, maximal characteristic distance and its energetic position are summarized in \autoref{t:rIC}.
			The solid line indicates the total characteristic distance in nanometres and relative to respective hydration shell radii $r_n$.
			The five strongest contributing capture channels are indicated in dashed lines. Their channel openings have been marked as well as some higher channel openings and the continuum threshold for assisted capture ($\infty$).
			The maximal allowed characteristic distance $\dist_{\max}$ of \watdmax{} is exclusively determined by the assisting water molecule.
}
\label{fig:rIC}
\end{figure*}
\begin{figure}[b!ht]
	\centering
	\includegraphics[width=0.5\textwidth]{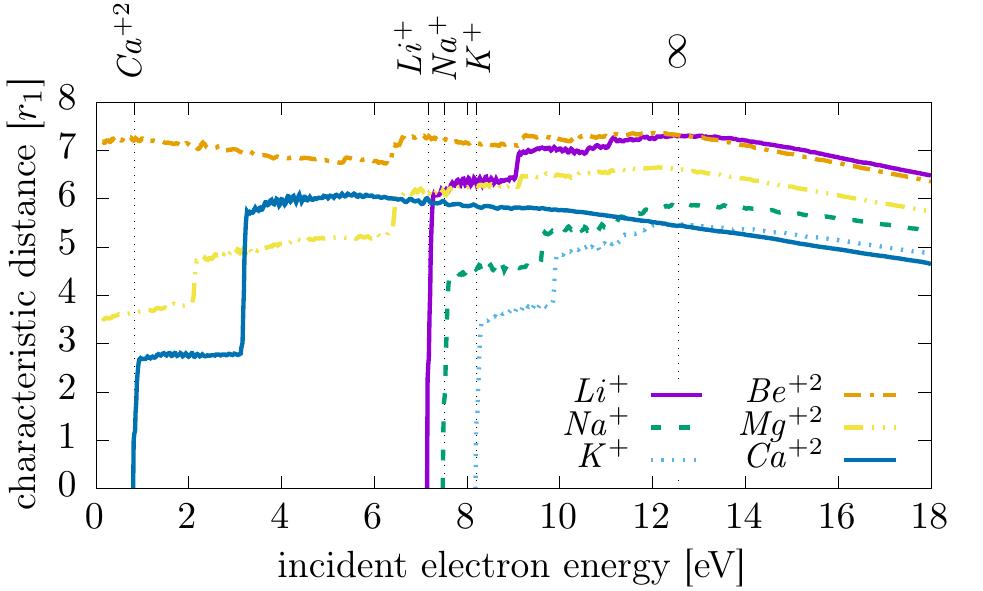}
	\caption{Characteristic distances $r_\IC$ of ICEC by alkali and alkaline earth metals in assistance by a water molecule in units of first solvation shell radius $r_1$.
	The calcium {dication} $Ca^{2+}$, for instance, has an energetic threshold of \Cathre{} for the incident free electron to allow an assisted electron capture by energy transfer to a water molecule. From this energetic onset, ICEC opens for capture into the $4s$ shell. The capture through assistance by water shows a strong gain in reach and is dominating over photorecombination up to $2.69~r_1$ above 0.95 eV from where it shows a plateau. As the additional assisted-capture channel into $3d$ opens at \CathreeD, the assisted-capture radius $\dist_{\IC}$ increases steeply with energy to reach up to $5.73~r_1$ at 3.26~eV. From there it presents a shallow increase with energy until reaching the maximum of $6.11~r_1$ at \Caemax\ from where it slowly descend with increasing energy. 
	It still reaches up to $4.65~r_1$ at 18~eV incident electron energy.
	}
	\label{fig:rIC/r1}
\end{figure}
The respective characteristic distances $\dist_{\IC}$ as function of incident electron energy $\epsilon$ are depicted in \autoref{fig:rIC} for assisted electron capture by alkali cations of lithium~(I) $\mathit{Li}^+$, sodium~(I) $\mathit{Na}^+$, and potassium~(I) $K^+$, as well as alkaline earth cations beryllium~(II) $\mathit{Be}^{2+}$, magnesium~(II) $\mathit{Mg}^{2+}$, and calcium~(II) $\mathit{Ca}^{2+}$ through ionisation of a water molecule. Key quantities are summarized in \autoref{t:rIC}. To compare the reach of environment assisted electron capture with respect to the respective dimension of hydration shells, the total ICEC radius is depicted in multiples of the first hydration shell radius in \autoref{fig:rIC/r1}.
{We stress that the ICEC radius represents a single-acceptor quantity. In a liquid environment, the contribution from each partner molecule adds to the total cross section. The interpretation of the ICEC radius in terms of the process's reach is therefore only a lower limit within the virtual photon approximation.}

\subsection{Alkali Monocations}\label{s:linak}
With an energetic threshold between \Lithre{}  ($Li^+$) and \Kthre{}  ($K^+$) to overcome before assisted electron capture opens, the characteristic distance $\dist_{\IC}$ shows a sharp onset between \LiSons{}  ($Li^+$) and \KSons{}  ($K^+$) for capture into the ground state $s$ shell with a clear step up to \LiPons{} for $Li^+$ and \KPons{} for $K^+$ at the threshold of the lowest $p$ shell. 
{Each of these opening onsets of assisted electron capture show clearly fine fluctuations within the first eV. This is the signature of molecular resonances in the water photoionisation cross section.}
The maximum of the characteristic distance between \Lidmax{}   ($Li^+$) to \Kdmax{}  ($K^+$) reaches very close to its captor-independent analytical limit $\dist_{\max}$ of 1.45~nm set solely by the assisting water molecule. Its position between \Liemax{} ($Li^+$) and \Kemax{} ($K^+$) coincides roughly with its respective continuum threshold energy.
Above this threshold, the individual contributions of capturing states as well as the total characteristic distance clearly wear the large-energy signature dictated by the photoionisation function of the assisting water molecule. While decreasing with increasing energy, the ICEC radius remains above \Lidend{} for energies up to 18~eV. This is {large} in comparison to the respective hydration shells. %

Intermolecular coulombic electron capture reaches thereby significantly beyond the third solvation shell for each alkali metal already from the opening of the channel.
The absolute assisted-capture radius is relatively similar across the elements but the first hydration shell grows with the atomic number. This leads to differences in the ICEC radius relative to the first solvation shell radius $r_1$.
At the opening plateau of assisted ground state capture ($2s$), it reaches beyond $6.1~r_1$ for lithium. It reaches $4.6~r_1$ for sodium at the opening plateau of assisted ground-state capture into $3s$, and still reaches beyond $3.5~r_1$ for the much larger potassium for the opening plateau of ground state capture into $4s$. \autoref{fig:rIC/r1} shows the respective total ICEC radius in units of first solvation shell radius $r_1$ for the investigated alkali monocations as well as alkaline
earth dications. With additional channels open at higher incident energies all three capturing cations show ICEC radii larger than $5.0~r_1$ up to 17~eV.

\subsection{Alkaline Earth Dications}\label{s:bemgca}
The dications of alkaline earth metals show an ICEC radius above 1~nm for a significantly larger energy range than alkali monocations, and more variation in their ICEC radius behaviour among each other owing to their larger charge and spread-out energetic thresholds of capture channels.
The difference between ionisation potentials of beryllium (II) and water is \Bethre{}. This indicates that the assisted capture is in fact already open for an incident electron with vanishing kinetic energy. The same holds for magnesium (II), while calcium (II) has an assisted-capture threshold at \Cathre{} which is close to zero but positive (cf. \autoref{t:rIC}).
As a result, $\mathit{Be}^{2+}$ already allows assisted capture into the $2s$ and $2p$ shells from vanishing incident energies which already presents the decaying tail of $\sigma_{\wat}/(h\nu)^4$ with increasing energy in the ICEC radius. This trend is interrupted by distinct steps upwards due to channel openings for capture into higher shells.
The magnesium (II) cation with its intermediate threshold of \Mgthre{} allows only $3s$ capture at vanishing incident energy, but the second assisted capture channel into $3p$ opens at 2.05~eV. This produces an overall increase of the ICEC radius with increasing incident energy up to its maximum of \Mgdmax{} at \Mgemax. 
The characteristic distance for assisted-capture by a calcium (II)  cation looks arguably most similar to that of the alkali monocations with comparison to the other two alkaline earth dications: It shows a positive energetic threshold for ground state capture, a significant stepwise increase with the opening of the second capture channel, here $3d$ capture at \CathreeD, %
and a global maximum reached closely thereafter, here at \Caemax. The absolute maximum ICEC radius of \Cadmax{} is closer to those reached by the alkali monocations and only 3.33\% short of the limiting $\dist_{\max}$ despite the curve's shift to lower incident energies.

\paragraph{}
With respect to the first hydration shell radius $r_1$, ICEC on alkaline earth dications has a large reach beyond $4.7~r_1$ for a range over more than 14.5~eV from energy thresholds of at least two open assisted capture channels upwards due to their higher charge and tighter hydration radius compared to alkali monocations.
While the alkaline earth dications fell short of the maximal allowed $\dist_{\max}$ in comparison of their absolute characteristic distance to their alkali monocationic counterparts, they are actually reaching further with respect to their respective hydration shell radius $r_1$ (cf. \autoref{fig:rIC/r1}):
The ICEC radius $\dist_{\IC}$ reaches up to $7.35~r_1$ for $Be^{2+}$, up to $6.65~r_1$ for $Mg^{2+}$, and up to $6.09~r_1$ for $Ca^{2+}$.
Particularly above 13~eV incident electron energy, the characteristic distances with respect to the first hydration shell radius are comparable for the pairs of $Li^+$ and $Be^{2+}$ as well as for $K^+$ and $Ca^{2+}$ and show the predicted high-energy tail behaviour of slow decay with increasing electron energy.
The assisted capture radius $\dist_{\IC}$ still reaches significantly beyond $4.5~r_1$ at 18~eV for all the investigated metal cations.

\section{Conclusions}\label{s:conclusion}
In this work, we introduced the characteristic distance $\dist_\IC$ for interparticle coulombic electron capture (ICEC) as a measure of quantum efficiency with respect to environment-independent photorecombination. In equivalence to the F\"orster radius for F\"orster resonant energy transfer, this $\dist_{\IC}$ allows to interpret the reach of ICEC.
We have furthermore presented experimentally relevant limits that can easily be evaluated to classify the significance of ICEC for any given pair of electron captor and assisting partner.
Notably, the ICEC radius as function of incident electron energy is mainly shaped by the photoionisation cross section of the assisting partner molecule.
The reach of ICEC was evaluated for bio-relevant alkali monocations $Li^+$, $Na^+$, $K^+$ and alkaline earth dications $Be^{2+}$, $Mg^{2+}$, and $Ca^{2+}$ by assistance from a water molecule.

Assisted capture of a free electron %
dominates significantly over en\-vi\-ron\-ment-in\-de\-pendent photo\-re\-com\-bi\-nation for dis\-tances between the reaction partners far beyond the third hydration shell radius. The assisted capture radius $\dist_{\IC}$ exceeded even distances 4 times that of the respective radius of the first hydration shell {$\dist_1$} for an energy range of at least 10~eV.
{
	The maximum reaches ranges between $5.5~\dist_1$  for $K^+$, and $7.4~\dist_1$ for $Be^{2+}$.
	Alkaline earth metal dications $Be^{2+}$, $Mg^{2+}$ and $Ca^{2+}$ are active at low incident energy. The ICEC reaction pathway is here already open at vanishing incident energy for the smaller dications and opens for calcium at \Cathre.
	The investigated alkali monocations $Li^+$, $Na^+$ and $K^+$ show a clear energetic threshold between \Lithre{} to \Kthre{} for the incident electron to be captured through the ICEC channel.
}

The introduced measures $\dist_{\IC}$ and $\dist_{\max}$ for the reach of ICEC will allow easy access to the design of future dedicated experimental measurements. Particularly the choice of the assisting partner molecule is essential. In other environmental contexts for instance, the introduced maximal value $\dist_{\max}$ can be quickly estimated. For instance, while it was 1.453~nm for an assisting water molecule, it would be 1.635~nm for an assisting ethanol molecule and 2.357~nm in the case of a carbon dioxide molecule as assisting partner. Species with a higher photoionisation cross section are therefore expected to assist ICEC over even longer distances.

The far reach of assisted electron capture has %
considerable implications for our understanding of reactions induced by slow electrons in any environment, but particularly in the context of propagating radiation damage in biological systems:
Slow secondary electrons induced by radiation damage in biological systems predominantly recombine with solvated cations through water-molecule-assisted capture rather than via photorecombination. This occurs within at least the first and second hydration shell when the threshold energy is met. 
It reduces the cation's bio-chemical availability. A tertiary electron emerges from the assisting water molecule with a different energy. While the number of free electrons remains constant during this process, the kinetic energy changes according to the difference in ionisation thresholds. 

\appendix

\printbibliography

@Misc{NIST_ASD,
	author = {A.~Kramida and {Yu.~Ralchenko} and
	J.~Reader and {and NIST ASD Team}},
	HOWPUBLISHED = {{NIST Atomic Spectra Database
	(ver. 5.10), [Online]. Available:
	{\tt{https://physics.nist.gov/asd}} [2022, November 25].
	National Institute of Standards and Technology,
	Gaithersburg, MD.}},
	year = {2022},
}

@article{bushaw-pra2007,
	title =  { Ionization energy of {$^{6,7}\mathrm{Li}$} determined by triple-resonance laser spectroscopy },
	author = { Bushaw, B. A. and N\"{o}rtersh\"{a}user, W. and Drake, G. W. F. and Kluge, H.-J. },
	journal = {Phys. Rev. A},
	volume = {75},
	issue = {5},
	pages = {052503},
	numpages = {8},
	year = {2007},
	month = {5},
	publisher = {American Physical Society},
	doi = {10.1103/PhysRevA.75.052503},
}

@article{LORENZEN1981370,
	title = {Precise measurements of {39K nS and nD} energy levels with an evaluated wavemeter},
	journal = {Optics Communications},
	volume = {39},
	number = {6},
	pages = {370-374},
	year = {1981},
	issn = {0030-4018},
	doi = {https://doi.org/10.1016/0030-4018(81)90225-X},
	url = {https://www.sciencedirect.com/science/article/pii/003040188190225X},
	author = {C.-J. Lorenzen and K. Niemax and L.R. Pendrill}
}

@article{C-JLorenzen_1983,
	doi = {10.1088/0031-8949/27/4/012},
	url = {https://dx.doi.org/10.1088/0031-8949/27/4/012},
	year = {1983},
	month = {4},
	publisher = {},
	volume = {27},
	number = {4},
	pages = {300},
	author = {C-J Lorenzen and  K Niemax},
	title = {Quantum Defects of the n2P1/2,3/2 Levels in 39K I and 85Rb I},
	journal = {Physica Scripta}
}

@article{7185EL,
	journal={J. Phys. Chem. Ref. Data},
	volume={14},
	suppl={2},
	pages={1--664},
	title={Atomic Energy Levels of the Iron-Period Elements: Potassium through Nickel},
	author={J. Sugar and C. Corliss},
	year={1985},
	notes={NIST compilation}
}

@article{doi:10.1063/1.555879,
	author = {Kaufman,V.  and Martin,W. C. },
	title = {Wavelengths and Energy Level Classifications of Magnesium Spectra for All Stages of Ionization (Mg I through Mg XII)},
	journal = {Journal of Physical and Chemical Reference Data},
	volume = {20},
	number = {1},
	pages = {83-152},
	year = {1991},
	doi = {10.1063/1.555879},
	URL = { 
	https://doi.org/10.1063/1.555879
	},
	eprint = { 
	https://doi.org/10.1063/1.555879
	}	
}

@article{AEKramida_2005,
	doi = {10.1238/Physica.Regular.072a00309},
	url = {https://dx.doi.org/10.1238/Physica.Regular.072a00309},
	year = {2005},
	month = {1},
	publisher = {},
	volume = {72},
	number = {4},
	pages = {309},
	author = {A E Kramida},
	title = {Critical Compilation of Wavelengths and Energy Levels of Singly Ionized Beryllium (Be II)},
	journal = {Physica Scripta}
}

@article{Zobeley01,
	author = {Zobeley, J. and Santra, R. and Cederbaum, L. S.},
	title = {{Electronic decay in weakly bound heteroclusters: Energy transfer versus electron transfer}},
	journal = {J. Chem. Phys.},
	publisher = {AIP},
	year = {2001},
	volume = {115},
	issue = {11},
	pages = {5076},
	numpages = {13},
	note = {theory},
	doi = {https://doi.org/10.1063/1.1395555}
}

@article{sisourat-pra2018,
	title = {Interatomic Coulombic electron capture from first principles},
	author = {Sisourat, Nicolas and Miteva, Tsveta and Gorfinkiel, Jimena D. and Gokhberg, Kirill and Cederbaum, Lorenz S.},
	journal = {Phys. Rev. A},
	volume = {98},
	issue = {2},
	pages = {020701},
	numpages = {4},
	year = {2018},
	month = {8},
	publisher = {American Physical Society},
	doi = {10.1103/PhysRevA.98.020701},
}

@article{molle-jcp-2019,
	author = {{Molle}, {A} and E R Berikaa and F M Pont and A Bande* },
	title = {Quantum Size Effect Affecting Environment Assisted Electron Capture in Quantum Confinements},
	journal = {Journal of Chemical Physics},
	publisher = {American Chemical Society},
	volume = {150},
	number = {22},
	pages = {224105},
	year = {2019},
	doi = {10.1063/1.5095999},
}

@article{pont-physrev-2013,
	author = {F M Pont and A Bande and L S Cederbaum},
	title = {Controlled energy-selected electron capture and release in double quantum dots},
	journal = {Phys. Rev. B},
	year = {2013},
	volume = {88},
	issuesubtitle = {Rapid Communications},
	pages = {241304(\textbf{R})},
	doi = {10.1103/PhysRevB.88.241304},
}

@Article{bande2013-04031,
	Title                    = {Dynamics of Interatomic Coulombic Decay in Quantum Dots: Singlet Initial State},
	Author                   = {Bande, A. and Pont, F. M. and Dolbundalchok, P. and Gokhberg, K. and Cederbaum, L. S.},
	Journal                  = {EPJ Web Conf.},
	Year                     = {2013},
	Pages                    = {04031},
	Volume                   = {41},
}

@online{icdrefbase,
	editor = {N. V. Kryzhevoi},
	subtitle = {{ICD refbase}},
	title = {Interatomic (Intermolecular) Coulombic Decay and Related Phenomena},
	date = {2020},
	url = {https://www.pci.uni-heidelberg.de/tc/usr/icd/ICD.refbase.html},
	urldate = {2022-11-22},
	archivePrefix = {icd.refbase.html}
}

@article{molle-pra2021,
	title = {Fano interferences in environment-enabled electron capture},
	author = { Molle, A and Dubois, Alain and Gorfinkiel, Jimena D. and Cederbaum, Lorenz S. and Sisourat*, Nicolas},
	journal = {Phys. Rev. A},
	volume = {103},
	issue = {1},
	pages = {012808},
	numpages = {5},
	year = {2021},
	month = {01},
	publisher = {American Physical Society},
	doi = {10.1103/PhysRevA.103.012808},
}

@article{molle-pra2021b,
	title = {Electron attachment to a proton in water by interatomic Coulombic electron capture: An $R$-matrix study},
	author = { Molle, A and Dubois, Alain and Gorfinkiel, Jimena D. and Cederbaum, Lorenz S. and Sisourat*, Nicolas},
	journal = {Phys. Rev. A},
	volume = {104},
	issue = {2},
	pages = {022818},
	numpages = {5},
	year = {2021},
	month = {08},
	publisher = {American Physical Society},
	doi = {10.1103/PhysRevA.104.022818},
}

@article{pont-jpcm-2019,
	doi = {10.1088/1361-648x/ab41a9},
	year = {2020},
	volume = {32},
	pages = {065302},
	publisher = {{IOP} Publishing},
	author = {Federico M. Pont and A Molle and Essam R. Berikaa and Sascha Bubeck and Annika Bande},
	title = {Predicting the performance of the inter-{Coulombic} electron capture from single-electron quantities},
	journal = {Journal of Physics: Condensed Matter},
}

@phdthesis{molle-dr-2019,
	author = {A Molle},
	title = {Electron Dynamics of Interatomic Coulombic Electron Capture in Artificial and Real Atoms},
	type = {Inaugural Dissertation to obtain the academic degree
	Doctor rerum naturalium (Dr.~rer.~nat.)},
	institution = {Theoretical Chemistry, Department of Biology, Chemistry and Pharmacy, Freie Universit\"at Berlin},
	date = {2019-11-13},
	nonote = {supervised by Dr Annika Bande and Prof Dr Beate Paulus},
	month = {11},
	year = {2019},
	doi = {10.17169/refubium-25683},
}

@article{molle-phd-2019,
	crossref = {molle-dr-2019},
	nonote = {Electron Dynamics of Interatomic Coulombic Electron Capture in Artificial and Real Atoms},
	journal = {{FU Dissertationen}},
	type = {Dissertation, Freie Universität Berlin University Library},
	noaddendum = {{Inaugural Dissertation to obtain the academic degree
	{\textit{Doctor rerum naturalium}} (Dr.~rer.~nat.)}, {supervised by Dr Annika Bande and Prof Dr Beate Paulus}},
	noeprint = {http://dx.doi.org/10.17169/refubium-25683},
	publisher = {Freie Universität Berlin University Library}
}

@article{gokhberg-jphysb2009,
	author = {Gokhberg, K and Cederbaum, L S},
	title = {Environment assisted electron capture},
	journaltitle = {J.~Phys.~B},
	date = {2009},
	annotejournalsubtitle = {At.~Mol.~Opt.~Phys.},
	issuesubtitle = {Fast Track Communication},
	volume = {42},
	pages = {231001},
	doi = {10.1088/0953-4075/42/23/231001},
}

@article{gokhberg-physrev2010,
	author = {K Gokhberg and L S Cederbaum},
	title = {Interatomic Coulombic electron capture},
	journaltitle = {Phys.~Rev.~A},
	date = {2010},
	volume = {82},
	pages = {052507},
	doi = {10.1103/PhysRevA.82.052707},
}

@inbook{oxenius1986,
	author = {Joachim Oxenius},
	chapter = {1.6.5 Photoionization and Radiative Recombination},
	booktitle = {Kinetic Theory of
	Particles and Photons},
	date = {1986},
	booksubtitle = {Theoretical Foundations of
	{Non-LTE} Plasma Spectroscopy },
	edition = {Softcover reprint of the hardcover 1st edition 1986},
	publisher = {Springer-Verlag},
	location = {Berlin Heidelberg},
	doi = {10.1007/978-3-642-70728-5},
	NOTEurl = {https://link.springer.com/content/pdf/10.1007%2F978-3-642-70728-5},
	NOTEurldate = {2020-07-09}
}

@article{zatsarinny-cpc2006,
	title = {{BSR}: {B-spline} atomic {R-matrix} codes},
	journal = {Computer Physics Communications},
	volume = {174},
	number = {4},
	pages = {273-356},
	year = {2006},
	issn = {0010-4655},
	doi = {10.1016/j.cpc.2005.10.006},
	author = {Oleg Zatsarinny}
}

@article{zatsarinny-physreva2008,
	title = {Relativistic {$B$-spline} {$R$-matrix} method for electron collisions with atoms and ions: Application to low-energy electron scattering from {Cs}},
	author = {Zatsarinny, Oleg and Bartschat, Klaus},
	journal = {Phys. Rev. A},
	volume = {77},
	issue = {6},
	pages = {062701},
	numpages = {7},
	year = {2008},
	month = {06},
	publisher = {American Physical Society},
	doi = {10.1103/PhysRevA.77.062701},
}

@article{loeffler-jcp2002,
	author = {Loeffler,Hannes H.  and Rode,Bernd M. },
	title = {The hydration structure of the lithium ion},
	journal = {The Journal of Chemical Physics},
	volume = {117},
	number = {1},
	pages = {110-117},
	year = {2002},
	doi = {10.1063/1.1480875},
}

@article{lyubartsev-jcp2001,
	author = {Lyubartsev,A. P.  and Laasonen,K.  and Laaksonen,A. },
	title = {Hydration of {Li\textsuperscript{+}} ion. An ab initio molecular dynamics simulation},
	journal = {The Journal of Chemical Physics},
	volume = {114},
	number = {7},
	pages = {3120-3126},
	year = {2001},
	doi = {10.1063/1.1342815},	
}

@article{smirnov-rjgc2008,
	author = {Smirnov, P. R. and Trostin, V. N.},
	date = {2008-09-01},
	title = {Structural parameters of hydration of {Be}\textsuperscript{2+} and {Mg}\textsuperscript{2+} ions in aqueous solutions of their salts},
	journal = {Russian Journal of General Chemistry},
	pages = {1643--1649},
	volume = {78},
	issue = {9},
	doi = {10.1134/S1070363208090016},
}

@article{rudolph-dalton2009,
author = {Rudolph, Wolfram W. and Fischer, Dieter and Irmer, Gert and Pye, Cory C.},
title  = {Hydration of beryllium(ii) in aqueous solutions of common inorganic salts. A combined vibrational spectroscopic and ab initio molecular orbital study},
journal  = {Dalton Trans.},
year  = {2009},
issue  = {33},
pages  = {6513-6527},
publisher  = {The Royal Society of Chemistry},
doi  = {10.1039/B902481F},
}

@article{howell-jpcm1996,
	doi = {10.1088/0953-8984/8/25/004},
	url = {https://doi.org/10.1088%2F0953-8984%2F8%2F25%2F004},
	year = 1996,
	month = {06},
	publisher = {{IOP} Publishing},
	volume = {8},
	number = {25},
	pages = {4455--4463},
	author = {I Howell and G W Neilson},
	title = {hydration in concentrated aqueous solution},
	journal = {Journal of Physics: Condensed Matter}
}

@article{kiyohara-jcp2019,
	author = {Kiyohara,Kenji  and Kawai,Yusuke },
	title = {Hydration of monovalent and divalent cations near a cathode surface},
	journal = {The Journal of Chemical Physics},
	volume = {151},
	number = {10},
	pages = {104704},
	year = {2019},
	doi = {10.1063/1.5113738},
}

@article{egorov-jml2000,
	title = "Influence of temperature on the microstructure of the lithium-ion hydration shell. A molecular dynamics description",
	journal = "Journal of Molecular Liquids",
	volume = "89",
	number = "1",
	pages = "47 - 55",
	year = "2000",
	issn = "0167-7322",
	doi = "https://doi.org/10.1016/S0167-7322(00)90004-7",
	author = "A.V. Egorov and A.V. Komolkin and V.I. Chizhik"
}

@article{galib-jcp2017,
	author = {Galib,M.  and Baer,M. D.  and Skinner,L. B.  and Mundy,C. J.  and Huthwelker,T.  and Schenter,G. K.  and Benmore,C. J.  and Govind,N.  and Fulton,J. L. },
	title = {Revisiting the hydration structure of aqueous {Na\textsuperscript{+}}},
	journal = {The Journal of Chemical Physics},
	volume = {146},
	number = {8},
	pages = {084504},
	year = {2017},
	doi = {10.1063/1.4975608},
}

@article{potts-prsl1972,
author = {{Potts}, A.~W. and {Price}, W.~C.},
title = "{Photoelectron Spectra and Valence Shell Orbital Structures of Groups V and VI Hydrides}",
journal = {Proceedings of the Royal Society of London Series A},
year = 1972,
month = jan,
volume = {326},
number = {1565},
pages = {181-197},
doi = {10.1098/rspa.1972.0004},
}

@article {fowler-pnas1925,
	author = {Fowler, R. H. and Milne, E. A.},
	title = {A Note on the Principle of Detailed Balancing},
	volume = {11},
	number = {7},
	pages = {400--402},
	year = {1925},
	doi = {10.1073/pnas.11.7.400},
	publisher = {National Academy of Sciences},
	issn = {0027-8424},
	NOTEurl = {https://www.pnas.org/content/11/7/400},
	NOTEeprint = {https://www.pnas.org/content/11/7/400.full.pdf},
	journal = {PNAS},
	NOTEjournal = {Proceedings of the National Academy of Sciences}
}

@article{heays-aa2017,
	author = {{Heays, A. N.} and {Bosman, A. D.} and {van Dishoeck, E. F.}},
	title = {Photodissociation and photoionisation of atoms and molecules   of astrophysical interest},
	DOI= "10.1051/0004-6361/201628742",
	url= "https://doi.org/10.1051/0004-6361/201628742",
	journal = {A\&A},
	year = 2017,
	volume = 602,
	pages = "A105",
}

@article{vanharrevelt-jpca2008,
	author = {{van Harrevelt}, Rob and {van Hemert}, Marc C.},
	title = {Quantum Mechanical Calculations for the {H\textsubscript2O + $h\nu$ \ensuremath{\to} O(\textsuperscript1D) + H\textsubscript2} Photodissociation Process},
	journal = {The Journal of Physical Chemistry A},
	volume = {112},
	number = {14},
	pages = {3002-3009},
	year = {2008},
	doi = {10.1021/jp711857w},
	note ={PMID: 18338878},	
}

@article{mordaunt-jcp1994,
	author = {Mordaunt, D. H.  and Ashfold, M. N. R.  and Dixon, R. N. },
	title = {Dissociation dynamics of {H\textsubscript2O(D\textsubscript2O)} following photoexcitation at the {Lyman-$\alpha$} wavelength (121.6 nm)},
	journal = {The Journal of Chemical Physics},
	volume = {100},
	number = {10},
	pages = {7360-7375},
	year = {1994},
	doi = {10.1063/1.466880},	
}

@article{smith-apj1981,
	author = {{Smith}, P.~L. and {Yoshino}, K. and {Griesinger}, H.~E. and {Black}, J.~H.},
	title = "{Oscillator strengths for lines of the F/0, 0, 0/-X/0, 0, 0/ band of H2O at 111.5 nanometers and the abundance of H2O in diffuse interstellar clouds}",
	journal = {Astrophysical Journal},
	abbrev = {Ap J},
	year = 1981,
	month = {11},
	volume = {250},
	pages = {166-174},
	doi = {10.1086/159359},
	adsurl = {https://ui.adsabs.harvard.edu/abs/1981ApJ...250..166S}
}

@article{chan-chemphys1993,
	title = "The electronic spectrum of water in the discrete and continuum regions. Absolute optical oscillator strengths for photoabsorption (6–200 eV)",
	journal = "Chemical Physics",
	volume = "178",
	number = "1",
	pages = "387 - 400",
	year = "1993",
	issn = "0301-0104",
	doi = "10.1016/0301-0104(93)85078-M",
	author = "W.F. Chan and G. Cooper and C.E. Brion"
}

@ARTICLE{verner-apj1996,
	author = {{Verner}, D.~A. and {Ferland}, G.~J. and {Korista}, K.~T. and
	{Yakovlev}, D.~G.},
	title = "{Atomic Data for Astrophysics. II. New Analytic FITS for Photoionization Cross Sections of Atoms and Ions}",
	journal = {Astrophysical Journal},
	abbrev = {Ap J},
	year = 1996,
	month = {07},
	volume = {465},
	pages = {487},
	doi = {10.1086/177435},
}

@article{mendoza-physcr1996,
	doi = {10.1088/0031-8949/1996/t65/030},
	year = 1996,
	month = {01},
	publisher = {{IOP} Publishing},
	volume = {T65},
	pages = {198--206},
	author = {C Mendoza},
	title = {Results from the opacity project and {TOPbase}},
	journal = {Physica Scripta}
}

@article{zatsarinny-physreva2010,
	title = {Photoionization of potassium atoms from the ground and excited states},
	author = {Zatsarinny, O. and Tayal, S. S.},
	journal = {Phys. Rev. A},
	volume = {81},
	issue = {4},
	pages = {043423},
	numpages = {9},
	year = {2010},
	month = {04},
	publisher = {American Physical Society},
	doi = {10.1103/PhysRevA.81.043423},
	url = {https://link.aps.org/doi/10.1103/PhysRevA.81.043423}
}

@article{yar-physreva2013,
	title = {Evidence of a {Cooper} minimum in the photoionization from the {$7s$} {$^{2}{S}_{1/2}$} excited state of potassium},
	author = {Yar, Ahmad and Ali, Raheel and Baig, M. Aslam},
	journal = {Phys. Rev. A},
	volume = {88},
	issue = {3},
	pages = {033405},
	numpages = {5},
	year = {2013},
	month = {09},
	publisher = {American Physical Society},
	doi = {10.1103/PhysRevA.88.033405},
}

@article{yoshino-chemphys1996,
	title = "Absorption cross section measurements of water vapor in the wavelength region 120 to 188 nm",
	journal = "Chemical Physics",
	volume = "211",
	number = "1",
	pages = "387 - 391",
	year = "1996",
	issn = "0301-0104",
	doi = "10.1016/0301-0104(96)00210-8",
	author = "K. Yoshino and J.R. Esmond and W.H. Parkinson and K. Ito and T. Matsui"
}

@article{yoshino-chemphys1997,
	title = "Absorption cross section measurements of water vapor in the wavelength region 120 nm to 188 nm (Chem. Phys. 211 (1996) 387-391)",
	journal = "Chemical Physics",
	volume = "215",
	number = "3",
	pages = "429 - 430",
	year = "1997",
	issn = "0301-0104",
	doi = "10.1016/S0301-0104(96)00381-3",
	author = "K. Yoshino and J.R. Esmond and W.H. Parkinson and K. Ito and T. Matsui"
}

@article{parkinson-chemphys2003,
	title = "Absorption cross-section measurements of water vapor in the wavelength region 181-199 nm",
	journal = "Chemical Physics",
	volume = "294",
	number = "1",
	pages = "31 - 35",
	year = "2003",
	issn = "0301-0104",
	doi = "10.1016/S0301-0104(03)00361-6",
	author = "W.H Parkinson and K Yoshino"
}

@article{fillion-jphysb2003,
	doi = {10.1088/0953-4075/36/13/308},
	year = 2003,
	month = {06},
	publisher = {{IOP} Publishing},
	volume = {36},
	number = {13},
	pages = {2767--2776},
	author = {J-H Fillion and F Dulieu and S Baouche and J-L Lemaire and H W Jochims and S Leach},
	title = {Ionization yield and absorption spectra reveal superexcited Rydberg state relaxation processes in H2O and D2O},
	journal = {Journal of Physics B: Atomic, Molecular and Optical Physics}
}

@article{fillion-jcp2004,
	author = {Fillion,J.-H.  and Ruiz,J.  and Yang,X.-F.  and Castillejo,M.  and Rostas,F.  and Lemaire,J.-L. },
	title = {High resolution photoabsorption and photofragment fluorescence spectroscopy of water between 10.9 and 12 eV},
	journal = {The Journal of Chemical Physics},
	volume = {120},
	number = {14},
	pages = {6531-6541},
	year = {2004},
	doi = {10.1063/1.1652566},	
}

@article{mota-chemphyslett2005,
	title = "Water VUV electronic state spectroscopy by synchrotron radiation",
	journal = "Chemical Physics Letters",
	volume = "416",
	number = "1",
	pages = "152 - 159",
	year = "2005",
	issn = "0009-2614",
	doi = "10.1016/j.cplett.2005.09.073",
	url = "http://www.sciencedirect.com/science/article/pii/S0009261405014429",
	author = "R. Mota and R. Parafita and A. Giuliani and M.-J. Hubin-Franskin and J.M.C. Lourenço and G. Garcia and S.V. Hoffmann and N.J. Mason and P.A. Ribeiro and M. Raposo and P. Limão-Vieira"
}

@article{foerster-andp1948,
	author = {F\"orster, Th.},
	title = {\foreignlanguage{german}{Zwischenmolekulare Energiewanderung und Fluoreszenz}},
	journal = {Annalen der Physik},
	year = {1948},
	volume = {437},
	number = {1-2},
	pages = {55-75},
	doi = {10.1002/andp.19484370105},
}

\section{Acknowledgements}
\
This project has received funding from the LabEx MiChem, part of French state funds managed by the ANR within the "Investissements d’Avenir" program under reference ANR-11-IDEX-0004-02.
The work of Axel Molle was supported in parts %
by the Research Foundation -- Flanders under FWO junior postdoctoral fellowship mandate %
1232922N.
Thomas-C. Jagau gratefully acknowledges funding from the European Research Council (ERC) under the European Union’s Horizon 2020 research and innovation program (Grant Agreement No. 851766).
The work of Oleg Zatsarinny was supported by the United States National Science Foundation under grant OAC-1834740 and the XSEDE allocation PHY-090031.

\section{Author Contribution Statement}
\
The mathematical derivation in this work was executed by Axel Molle.

The numerical data of atomic and molecular photoionisation cross sections used as input was acquired from published and openly available databases as indicated. The atomic data for the potassium monocation $K^+$ was computed and provided by Oleg Zatsarinny.
The data set for assisted electron-capture was produced by Axel Molle from the atomic and molecular input data.
Numerical data analysis and its graphical representation was undertaken by Axel Molle, so was the visual representation for the graphical abstract.

Primary textual composition was performed by Axel Molle.
The current version has undergone textual revision by Axel Molle and Nicolas Sisourat.
A previous version has seen redactoral revision by Thomas-C. Jagau and Nicolas Sisourat.
Alain Dubois, Nicolas Sisourat and Thoms-C. Jagau acted throughout this project in editorial capacity.

\end{document}